\documentclass[english,aps,prX,showpacs,twocolumn]{revtex4-1}
\usepackage[T1]{fontenc}
\usepackage[latin9]{inputenc}
\usepackage{babel}
\usepackage{graphicx}
\usepackage{subcaption}
\usepackage{bm}
\usepackage{bbm}
\usepackage{array}
\usepackage{amsmath}
\usepackage{amssymb}
\usepackage{dcolumn}
\usepackage{epstopdf}
\usepackage{bbold}
\usepackage{tikz}
\usepackage{pgfplots}
\usepackage[margin=15mm]{geometry}
\colorlet{linkequation}{blue}
\usepackage[colorlinks=true,linkcolor=blue,citecolor=blue,urlcolor=blue]{hyperref}
\usetikzlibrary{shadings}
\usepackage{siunitx}
\usepackage[nohyperlinks, nolist]{acronym}
\usepackage{braket}
\usepackage{mathrsfs}
\usepackage{import}
\usepackage{filecontents}
\usepackage{color}

\renewcommand{\vec}[1]{\bm{#1}}

\newcommand{\DT}[1]{\mathcal{D}_{#1}}
\newcommand{\GYR}{\mathcal{G}}
\newcommand{\MUS}{\mu_{\mathrm{s}}}
\newcommand{\PARTIAL}[2]{\dfrac{\partial {#1}}{\partial {#2}}}
\newcommand{\SPIN}{\vec{S}_i}

\newcommand{\HEFF}{\vec{H}_i}
\newcommand{\Jij}{\mathcal{J}_{ij}}

\newcommand{\TESLA}{\si{\tesla}}

\newcommand{\Q}{\mathrm{Q}}

\begin{document}

\title{Orientation dependent current-induced motion of skyrmions with various topologies
% in an ultrathin film
}

\author{Markus Wei\ss enhofer}
\email[]{markus.weissenhofer@uni-konstanz.de}
\affiliation{Fachbereich Physik, Universit\"at Konstanz, DE-78457 Konstanz, Germany}

\author{Ulrich Nowak}
\affiliation{Fachbereich Physik, Universit\"at Konstanz, DE-78457 Konstanz, Germany}

\date{\today}

%------------------------------------------------------------------------
\begin{abstract}
We study the current-driven motion of metastable localized spin structures with various topological charges in a (Pt\textsubscript{1-$x$}Ir\textsubscript{$x$})/Fe bilayer on a Pd(111) surface by combining atomistic spin model simulations with an approach based on the generalized Thiele equation. We demonstrate that besides a distinct dependence on the topological charge itself the dynamic response of skyrmionic structures with topological charges $\Q = -1$ and $\Q = 3$ to a spin-polarized current exhibits an orientation dependence. We further show that such an orientation dependence can be induced by applying an in-plane external field, possibly opening up a new pathway to the manipulation of skyrmion dynamics.
\end{abstract}

%-------------------------------------------------------------------------

%\keywords{..., }

\maketitle

\begin{acronym}
\acro{DMI}[DMI]{Dzyaloshinsky-Moriya interaction}
\acro{sLLG}[sLLG]{stochastic Landau-Lifshitz-Gilbert}
\acro{LLG}[LLG]{Landau-Lifshitz-Gilbert}
\acro{MSD}[MSD]{mean square displacement}
\acro{BP}[BP]{Belavin-Polyakov}
\end{acronym}

\section{Introduction}

Magnetic skyrmions are topologically protected spin configurations where the directions of the magnetic moments span the whole unit sphere \cite{Nagaosa2013}. As they are small in size and easily movable by applied currents, which can be significantly lower than those neccessary to trigger magnetic domain wall motion \cite{Jonietz2010,Yu2012,Schulz2012}, skyrmions are promising candidates for next generation magnetic logic and memory devices \cite{Fert2013,Iwasaki2013a,Zhou2014}.
It was shown by Belavin and Polyakov \cite{Belavin1975} that skyrmions exist as metastable excitations in the two-dimensional Heisenberg model. However, the presence of a finite magnetic anisotropy leads to a collapse of the skyrmion, hence an additional interaction term in the Hamiltonian is required in order to stabilize the radius of magnetic skyrmions. Several mechanisms for such an interaction have been identified so far: the \ac{DMI} \cite{DZYALOSHINSKY1958,Moriya1960,BOGDANOV1994255}, four-spin interactions \cite{Heinze2011} as well as the frustration of the Heisenberg-type exchange interactions \cite{Okubo2012}.
Past investigations have mostly focused on materials with \ac{DMI}, which requires inversion symmetry to be broken. A skyrmion lattice  was first detected in MnSi \cite{Muehlbauer2009} and since then, skyrmions have been found experimentally in a variety of other bulk systems \cite{Yu2010, Wilhelm2011,Muenzer2010,Adams2012,Kezsmarki2015,Tokunaga2015} as well as thin films \cite{Romming2013,Hsu2016,Moreau-Luchaire2016}. 

In recent theoretical works \cite{Rozsa2016,Rozsa2017}, the interplay between frustrated isotropic exchange interaction and \ac{DMI} has been identified as responsible for the formation and stability of isolated skyrmionic structures of various types in a (Pt\textsubscript{1-$x$}Ir\textsubscript{$x$})/Fe bilayer on Pd(111). These skyrmionic structures are characterized by their topological charge $\Q = 1/4\pi \int \mathrm{d}^2 r\ \vec{S} \cdot (\partial_x \vec{S} \times \partial_y \vec{S})$ with $\vec{S}$ being the unit vector describing the direction of the magnetic moment. It turned out that the skyrmion with $Q=1$ is energetically most favorable and an oscillating skyrmion-skyrmion interaction potential was found, which results from a competition between ferro- and antiferromagnetic isotropic exchange interactions. However, skyrmionic structures with other topological charges ranging from -2 to 3 were also found as metastable excitations. Although usually only the spin structure with $\Q=1$ is called skyrmion, for simplicity in the following all localized, metastable spin structures with integer topological charge will be referred to as skyrmions. 

Theoretical studies predict that skyrmions propagate along trajectories which deviate from the current direction \cite{Nagaosa2013,Zang2011,Iwasaki2013a,Iwasaki2013b,EverschorSitte2014,Mueller2015}, according to the so-called skyrmion Hall effect which has been experimentally demonstrated in thin film systems \cite{Jiang2016,Litzius2016}. This effect is a consequence of finite topological charge and therefore a viable method to experimentally distinguish between skyrmions and topologically trivial structures.

Recently, it was discovered that non-circular skyrmions entail a new pathway for obtaining certain functionalities as this feature introduces a new degree of freedom,  namely the orientation of the skyrmion. In numerical investigations, the anisotropy of the skyrmion profile was introduced via an in-plane magnetic field \cite{Lin2015} or via an anisotropic \ac{DMI}, either in an effort to model the effect of uniaxial stress \cite{Chen2018} or in a general investigation of the dependence of the skyrmion profile on the \ac{DMI} strength along different crystallographic axes \cite{Huang2017}. Experimentally, such non-circular skyrmions were observed in strained FeGe crystals \cite{Shibata2015}, where an anisotropic strain of $0.3\%$ induces distortions of the skyrmion lattice of the order of $20\%$.

In this paper, we discuss the current-driven motion of skyrmions with various topologies in the model system mentioned above, a (Pt\textsubscript{0.95}Ir\textsubscript{0.05})/Fe bilayer system on a Pd(111) surface.
We perform atomistic spin simulations based on the \ac{LLG} equation \cite{Landau1935,Gilbert2004} using a model Hamiltonian for the system parametrized by \textit{ab initio} calculations in Ref. \cite{Rozsa2016}. We demonstrate that the occurrence of non-circular skyrmions is intrinsic to that system and that the skyrmion Hall effect which appears during the current-induced skyrmion motion strongly depends on the respective topological charge as well as the shape and orientation of the skyrmion. The results are well supported by an approach based on the Thiele equation.

\section{Methods} %%%%%%%%%%%%%%%%%%%%%%%%%%%%%%%%%%%%%%%%%%%%%%
We model a (Pt\textsubscript{0.95}Ir\textsubscript{0.05})/Fe bilayer on a Pd(111) surface via a spin Hamiltonian $\mathcal{H}$ considering only the magnetic  Fe moments, 
\begin{equation}
\mathcal{H}
= \dfrac{1}{2} \sum_{i \neq j} \SPIN   \Jij \vec{S}_j 
+ \sum_i \SPIN \mathcal{K} \SPIN 
- \sum_i \MUS \SPIN \cdot \vec{B}.
\end{equation}

\begin{figure}
\centering
\includegraphics[scale=0.4]{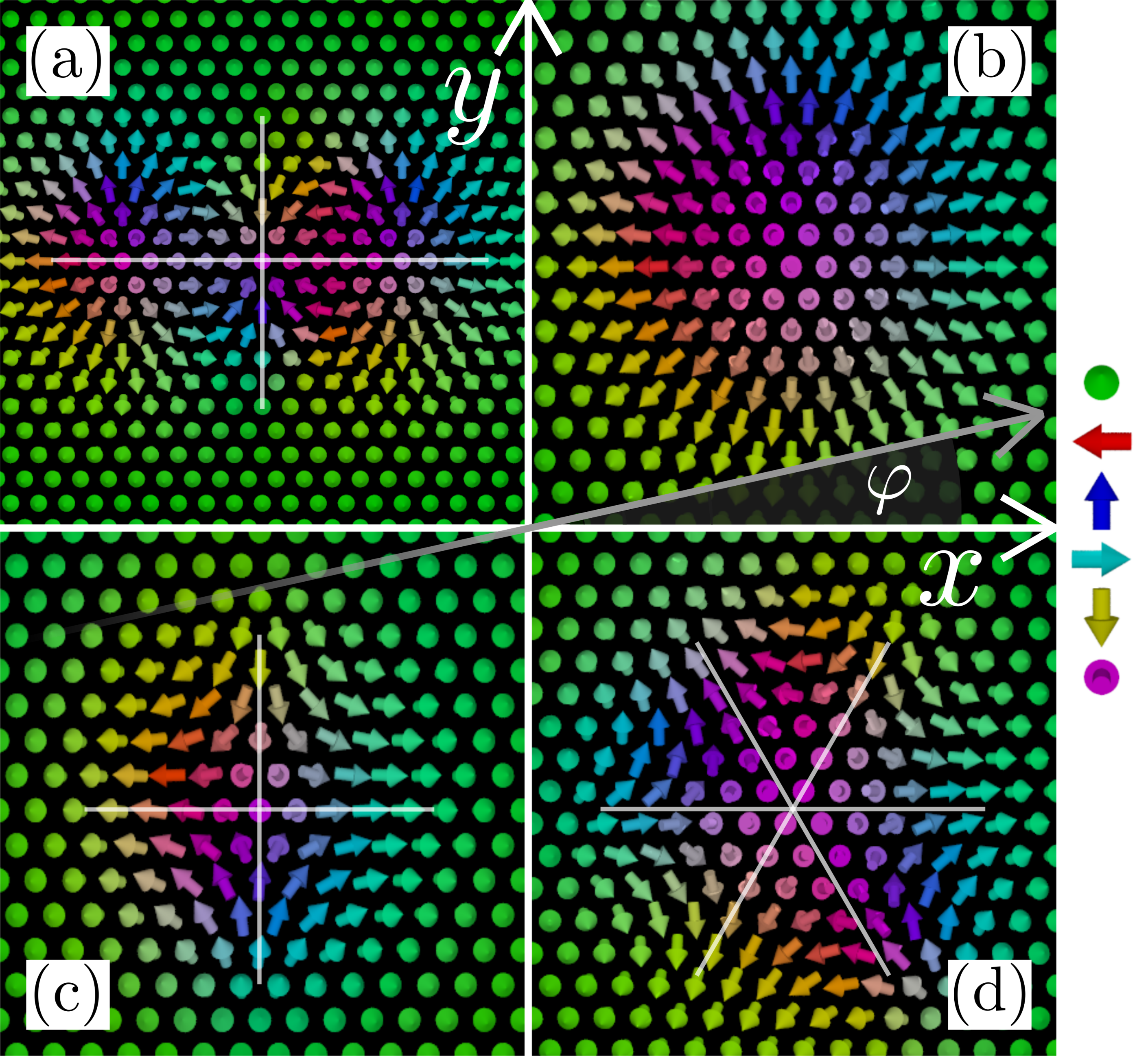}
\caption{Metastable spin configurations with different topological charges: skyrmions with (a) $\Q = 3$, (b) $\Q=1$, (c) $\Q=-1$, and (d) $\Q=-2$. The colors indicate the orientation of the spin vectors as illustrated on the right side of the figure. The grey arrow indicates the direction of the spin-polarized current that encloses the angle $\varphi$ with the $x$-axis. $B_{\perp}=2.35\TESLA$ in (a) and $0.5\TESLA$ in (b)-(d).}
\label{fig:skyrmion_types}
\end{figure}

The magnetic spin moment $\MUS$, the tensorial exchange coefficients $\Jij$ and the on-site anisotropy tensor $\mathcal{K}$ were taken from Ref. \cite{Rozsa2016,Zazvorka2019}, where they were determined from \textit{ab initio} calculations based on the screend Korringa-Kohn-Rostoker method \cite{Szunyogh1994,Szunyogh1995} and the relativistic torque method \cite{Szunyogh2003}. $\vec{B}$ is an external magnetic field applied perpendicular to the surface with a fixed value of $B_{\perp} = 0.5 \TESLA$ in the following, unless stated otherwise.
The ground state of the system is a spin spiral state which transforms into a collinear state when a field of $B_{\perp} \geqslant 0.21$ is applied \cite{Rozsa2016}. Skyrmionic spin structures with various topological charges $\Q$ can then occur as metastable excitations (skyrmions with $\Q=3,1-1,-2$ are depicted in Fig. \ref{fig:skyrmion_types}). While the \ac{DMI}, the antisymmetric part of the interaction tensor $\Jij$, present in this system renders skyrmions with $\Q=1$ energetically favorable, those with other topological charges are stabilized by the frustration of the exchange interaction, although being deformed by the \ac{DMI} \cite{Rozsa2017}. Solely the skyrmion with $Q=1$ keeps its circular shape, while the other textures become non-circular.\\
Furthermore it was found that in this system the \ac{DMI} induces a rotational pinning potential, as domain walls along different crystallographic axes possess different energies. This leads to preferred orientations of skyrmions with $\Q \neq 1$ with respect to the underlying lattice. Due to the symmetry of the system these preferred orientations possess a $C_{\mathrm{3v}}$ symmetry which means that rotating the spin structures by $2 \pi/3$ leads to equivalent configurations. We note that skyrmions with $\Q = 3$ and $\Q=-1$ have an additional $C_{\mathrm{2}}$ symmetry which in combination with the symmetry of the lattice lead to a $C_{\mathrm{6}}$ symmetry of the rotational pinning potential.\\

The dynamics of the spins is calculated by the means of the \ac{LLG} equation of motion  \cite{Nowak2007},

\begin{equation}\label{eq:sLLG}
\PARTIAL{\SPIN}{t} = - \dfrac{\gamma}{(1+\alpha^2)\MUS}\SPIN \times 
(\HEFF + \alpha \SPIN \times \HEFF).
\end{equation}

\noindent Here, $\alpha$ is the Gilbert damping parameter and $\gamma$ the absolute value of the gyromagnetic ratio. The effective field $\HEFF= -\partial \mathcal{H}/\partial \SPIN$ is generated by the interaction with the neighboring spins, the on-site anisotropy, and the external field. In order to include the effect of spin-polarized currents on spin textures such as skyrmions, the right-hand side of Eq (\ref{eq:sLLG}) is supplemented with a spin torque term $\vec{T}_{\mathrm{ST}}$ which consists of an adiabatic and a non-adiabatic contribution \cite{Zhang2004},
\begin{align}\label{eq:LLG_current} 
\vec{T}_{\mathrm{ST}}
& = \dfrac{1+\beta \alpha}{1+\alpha^2} \SPIN \times (\SPIN \times (\vec{u} \cdot \nabla )\SPIN )\nonumber \\
& + \dfrac{\beta-\alpha}{1+\alpha^2} \SPIN \times (\vec{u} \cdot \nabla )\SPIN,
\end{align}
with a constant non-adiabaticity parameter $\beta$ and the (spin-) drift velocity $\vec{u}$ of the conduction electrons proportional to the current density.\par 
The simulations are based on a GPU-accelerated numerical integration of the \ac{LLG} equation with the aditional spin torque term as described above using Heun's method \cite{Nowak2007} with a time step $\Delta t = \SI{1.4e-15}{\second}$. We simulate $96 \times 96$ spins featuring open boundary conditions 
with a Gilbert damping parameter $\alpha=0.1$ and a non-adiabaticity parameter $\beta=0.3$. \par
The numerical results are discussed in the light of theory using an approach proposed by Thiele \cite{Thiele1973} which allows for the derivation of an effective equation of motion for a rigid skyrmionic spin structure in a two-dimensional film with constant velocity $\vec{v}$ under the effect of a spin-polarized current from the \ac{LLG} equation. One obtains \cite{Thiaville2005}

\begin{equation}\label{eq:ThieleEq_current}
\vec{G} \times (\vec{v} - \vec{u}) + \mathcal{D} (\alpha \vec{v} - \beta \vec{u})  + \vec{F} = 0.
\end{equation}

\noindent where $\vec{F}$ is the external force acting on the skyrmion, $\vec{G} = \GYR \vec{e}_{\perp}$ the gyrocoupling vector and $\mathcal{D}$ the dissipation tensor. $\GYR$ and the entries of the dissipation tensor $\DT{\mu \nu}$ are given by

\begin{align}
\label{eq:GYR_val}
\GYR & = 
4 \pi \MUS \int \mathrm{d}^2 r\ \rho_{top}(\vec{r}) \\
\rho_{top}(\vec{r})& =
 \dfrac{1}{4 \pi} \vec{S} \cdot (\partial_x \vec{S} \times \partial_y \vec{S}) \\
\label{eq:DT_entries}
\DT{\mu \nu} & = 
\MUS \int \mathrm{d}^2 r\ \partial_\mu \vec{S} \cdot \partial_\nu \vec{S},
\end{align}
where $\rho_{top}(\vec{r})$ is the topological charge density and $\mu$ and $\nu$ are Cartesian components. The dissipation tensor is obviously symmetric with $\DT{xy} = \DT{yx}$. Moreover, for axissymmetric skyrmions for which at least one symmetry axis is aligned along the $x$- or $y$-axis, the off-diagonal terms in $\mathcal{D}$ are zero, which follows from the symmetry of the integrand in relation (\ref{eq:DT_entries}). For the isotropic $Q=1$ skyrmion, additionally, the diagonal elements are equal,  $\DT{xx} = \DT{yy}$.

The shape of the skyrmion profile can most readily be characterized via the following tensor $\Sigma$ whose elements, in analogy to the gyration tensor for a continuous distribution \cite{shimada1981statistical,vsolc1971shape}, are given via
\begin{align}
\Sigma_{\mu \nu} = \dfrac{\int \mathrm{d}^2 r\ \rho_{top}(\vec{r}) (r_{\mu}-R_{\mu})(r_{\nu}-R_{\nu})}{\int \mathrm{d}^2 r\ \rho_{top}(\vec{r})}
\end{align}
where $\vec{R}$ denotes the center of the topological charge density. As $\Sigma$ is real and symmetric by definition, it can be diagonalized with the real eigenvalues $E_1$ and $E_2$. Without loss of generality we assume $E_1 \geqslant E_\mathrm{2}$ in the following. The corresponding eigenvectors $e_1$ and $e_2$ are mutually orthogonal and point along the principal axes of an ellipsoid approximating the topological profile of the skyrmion. The aspect ratio $\sigma$ of that ellipsoid is given via $\sigma=\sqrt{E_1}/\sqrt{E_2}$ and its radius of gyration $R_{\mathrm{gyr}}$ via  $R_{\mathrm{gyr}}^2= \mathrm{Tr}(\Sigma) = E_1+E_2$ \cite{shimada1981statistical}.

In the absence of external forces ($\vec{F} = 0$), the velocity $\vec{v}$ for skyrmions with a fixed orientation is derived from Eq. (\ref{eq:ThieleEq_current}) by a simple matrix inversion as

\begin{equation}
\vec{v} = 
\dfrac{\vec{u}}{\alpha^2 \DT{xx} \DT{yy} + \GYR^2 } 
\begin{pmatrix}
\alpha \beta \DT{xx} \DT{yy} + \GYR^2 & (\beta-\alpha)\DT{yy}\GYR  \\
 -(\beta-\alpha)\DT{xx}\GYR  &\alpha \beta \DT{xx} \DT{yy} + \GYR^2 
\end{pmatrix}
\,
\end{equation}
which allows for an angular dependent evaluation by using polar coordinates, $\vec{u} = (u \cos \varphi, u \sin \varphi)$.
Analogous to the context of skyrmions in strained chiral magnets \cite{Chen2018} the resulting skyrmion velocities parallel and perpendicular to the current direction are given by 
\begin{align}\label{eq:v_parallel}
v_{\parallel} &= \dfrac{u \big(\alpha \beta \DT{xx} \DT{yy} +\GYR^2 
+ \cos\varphi \sin\varphi(\beta-\alpha) (\DT{yy}-\DT{xx})\GYR \big)}{\alpha^2 \DT{xx} \DT{yy}  + \GYR^2 }, \\
\label{eq:v_perp}
v_{\perp} &= \dfrac{u \GYR (\alpha - \beta)}{\alpha^2 \DT{xx} \DT{yy} + \GYR^2 }
\big(\DT{xx} + (\DT{yy}-\DT{xx})\sin^2\varphi \big).
\end{align}
Relation (\ref{eq:v_parallel}) shows already that the universal relation $v_{\parallel} \approx u$ indepedent of $\beta$ for realistic small $\alpha$ \cite{Iwasaki2013b} does not hold for $\DT{xx} \neq \DT{yy}$ since it depends on $\varphi$ then.
For $\DT{xx} = \DT{yy}$ however, $v_{\parallel}$ and $v_{\perp}$ are independent of $\varphi$.
The skyrmion Hall angle can straightforwardly be calculated via 
\begin{equation}\label{eq:Theta}
\Theta = \tan^{-1}(v_{\perp}/v_{\parallel})
\end{equation}
whose sign turns out to be equal to the sign of $\GYR (\alpha - \beta)$. Furthermore it follows that $\Theta(\varphi + \pi) = \Theta(\varphi)$ (which is equivalent to switching the sign of $u$) and therefore $\varphi$ is  between $0$ and $\pi$. 
 
\begin{table}
\begin{tabular}{ p{1.5cm} p{1.8cm} p{1.8cm} p{1.8cm} p{1.8cm}}
\hline 
\hline\noalign{\smallskip}
\mbox{}\phantom{-}$\Q$ & $\GYR/4\pi\MUS$ &  $\DT{xx}/4\pi\MUS$ &  $\DT{yy}/4\pi\MUS$ & $\phantom{-}\sigma$ \\ 
\hline\noalign{\smallskip}
	$\phantom{-}3$ & $ \phantom{-}3.013$ & $\phantom{-}3.239$ & $ \phantom{-}3.355$ & $1.707$\\
	$\phantom{-}1$ & $\phantom{-}1.087$ & $\phantom{-}1.625$ & $\phantom{-}1.625$ & $1.004$ \\
  	$-1$ & $-1.026$ & $\phantom{-}1.571$ & $\phantom{-}1.111$ & $1.601$\\  
	$-2$ & $-2.070$ & $\phantom{-}2.336$ & $\phantom{-}2.334$ & $1.008$\\
\hline
\hline
\end{tabular}
\caption{Gyrocoupling $\GYR$, dissipative tensor elements $\DT{xx}$ and $\DT{yy}$, and aspect ratio $\sigma$ calculated numerically for different types of skyrmions at rest and oriented as shown in Fig. \ref{fig:skyrmion_types}. $B_{\perp}=2.35\TESLA$ for $\Q=3$ and $0.5\TESLA$ for the others.}
\label{table:DT_discrete_anisotropic}
\end{table} 

\section{Results} %%%%%%%%%%%%%%%%%%%%%%%%%%%%%%%%%%%%%%%%%%%%

For the evaluation of the analytical formulae above we use numerically calculated values of $\GYR$, $\DT{xx}$ and $\DT{yy}$. These values for skyrmions at rest and oriented as shown in Fig. \ref{fig:skyrmion_types} and the respective values for $\sigma$ are summarized in Table \ref{table:DT_discrete_anisotropic}.
We find isotropic dissipation tensors ($\DT{xx} = \DT{yy}$) for skyrmions with $\Q=1$ and surprisingly also for $\Q=-2$ while they are anisotropic ($\DT{xx} \neq \DT{yy}$ ) for $\Q=3,-1$. The latter fact can be explained using similar arguments as in Ref. \cite{Rozsa2017} and can be traced back to the effect of the \ac{DMI} on cycloidal spin spirals with different helicities. Favoring the right-handed rotation, the \ac{DMI} leads to a distortion of left-handed spin spirals of spin structures thus causing a broadening of compact formations such as stripe domains or skyrmions with left-handed segments. Any cross section of the skyrmion with $\Q=1$ is right-handed meaning that its spin configuration is cylindrically symmetric.
The handedness of the cross section of $\Q=-1$ skyrmions is angular dependent: being oriented as depicted in Fig. \ref{fig:skyrmion_types} the spins undergo a right-handed rotation in $x$-direction and a left-handed rotation in $y$-direction, leading to an elongation of the skyrmion along the latter axis which is reflected by the different diagonal elements of the dissipation tensor.
For $\Q=-2$ skyrmions, any cross-section consists of one right- and one left-rotating segment giving rise to an isotropic deformation and consequently to an isotropic dissipation tensor.
The skyrmion with $\Q=3$ has one strictly right-rotating cross section along the short axis and a combination of right- and left-handed rotation along the long axis. However, although being highly elongated ($\sigma \approx 1.7)$ the dissipation tensor reveals only a slight anisotropy ($\DT{xx}/\DT{yy} \approx 0.96$). This leads to the conclusion that the shape of the topological profile of a skyrmion does not necessarily reflect its dynamic properties.

\begin{figure}
\begin{subfigure}{0.5\textwidth}

\subcaption{}
\includegraphics[scale=0.4,trim=0cm 0.0cm 0cm 1.0cm]{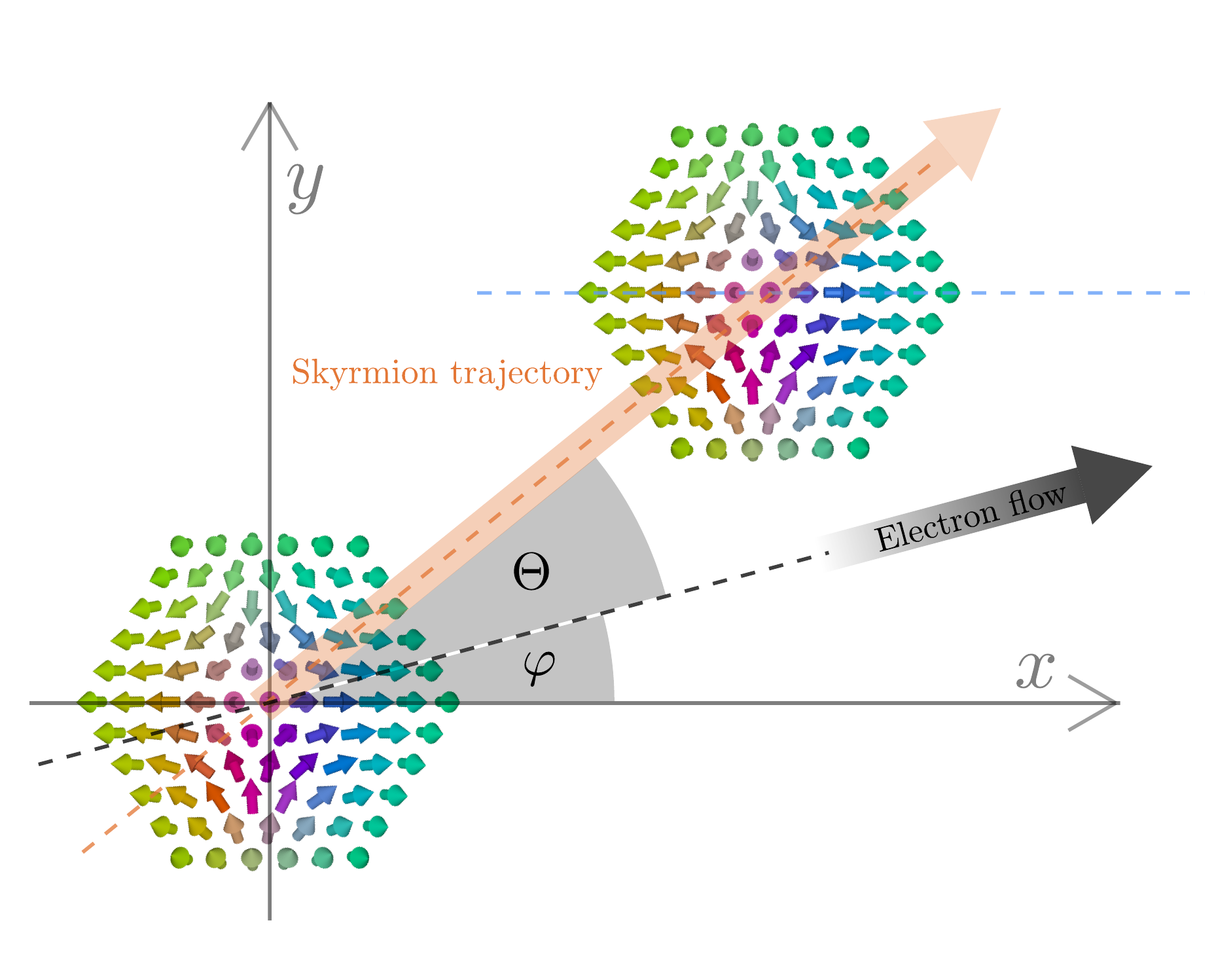}
\label{fig:SKH_sketch}
\end{subfigure}
\begin{subfigure}{0.5\textwidth}

\subcaption{}
\includegraphics[scale=0.9,trim=0cm 0cm 0cm 1.0cm]{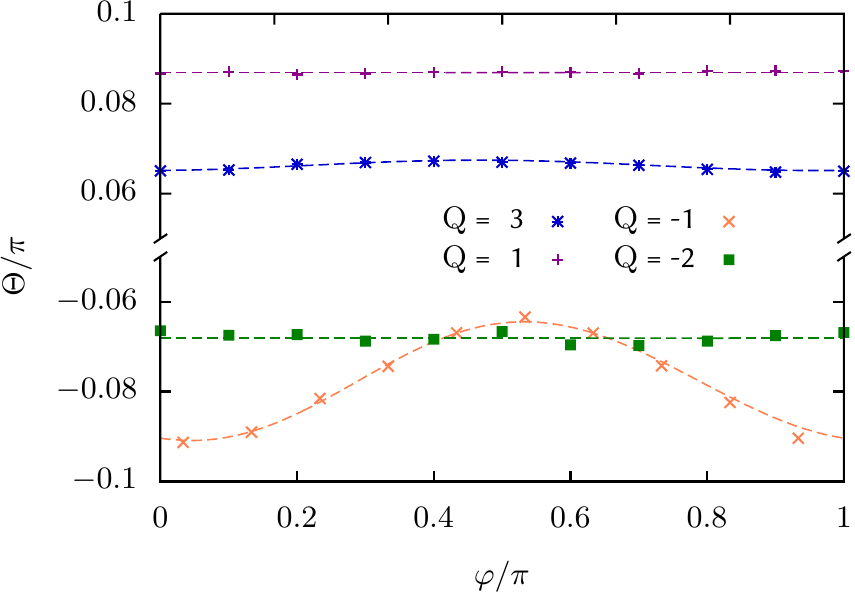}
\label{fig:SKHa_vs_phi}
\end{subfigure}

\caption{(a) Scheme of the skyrmion motion with definitions of angles $\Theta$ and $\varphi$. (b) Skyrmion Hall angle $\Theta$ versus angle of applied current $\varphi$ for skyrmions with topological charges $\Q$ as labeled. 
%$\alpha=0.1$ and $\beta=0.3$. 
$B_{\perp}=2.35\TESLA$ for $\Q=3$ and $0.5\TESLA$ otherwise. Dashed lines correspond to our semi-analytical calculations following Eq.\ \ref{eq:Theta}.}

\end{figure}

As the orientation of skyrmions with $\Q \neq 1$ is trapped due to the rotational pinning potential mentioned above \cite{Rozsa2017},  one can determine the orientation dependence of the Hall angle $\Theta(\varphi)$ in our simulations simply by varying the direction of the spin-polarized current with respect to the crystallographic axes (see Fig. \ref{fig:SKH_sketch}). Both, the simulations and the semi-analytic results based on relations (\ref{eq:v_parallel}),(\ref{eq:v_perp}) and (\ref{eq:Theta}) for $\Theta(\varphi)$ for skyrmions with $\Q=3,1,-2,-1$ are depicted in Fig. \ref{fig:SKHa_vs_phi}. In good agreement between theory and simulations we find a strong orientation dependence of the Hall angle for skyrmions with $Q=-1$, a weak dependence for $Q=3$, and hardly any dependence for $Q=1$ and -2. These findings are well reflected by the inequality (or equality) of the diagonal elements of the dissipative tensor (see Table \ref{table:DT_discrete_anisotropic}). The functional form of $\Theta(\varphi)$ is mainly determined by the $\sin^2 \varphi$-term in relation (\ref{eq:v_perp}) for $v_{\perp}$ but gets shifted a little due to the angular dependence of $v_{\parallel}$.

As $\alpha$ and $\beta$ are fixed parameters in a given system and because $v_{\parallel}$ and $v_{\perp}$ are both linear in the current density (in systems without pinning, see Ref. \cite{Litzius2016,Jiang2016}) the most accessible path to manipulate the skyrmion Hall angle is by manipulating the shape and size of skyrmions. This can easily be achieved via the applied external field $\vec{B}$, either by varying the strength of the out-of-plane component or by applying an additional in-plane field $B_{\parallel}$ \cite{Lin2015}.

\begin{figure}
\includegraphics[scale=1.0]{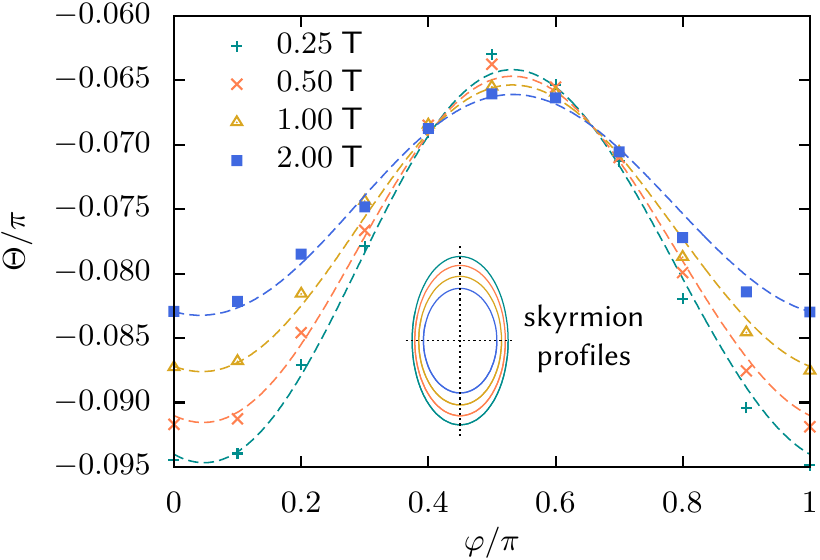}
\caption{Skyrmion Hall angle $\Theta$ versus direction of applied current $\varphi$ for skyrmions with $\Q=-1$
% with $\alpha=0.1$ and $\beta=0.3$ 
for different strengths of the out-of-plane field $B_{\perp}$. Dashed lines correspond to our semi-analytical calculations following Eq.\ \ref{eq:Theta}. Also shown are the topological skyrmion profiles. Their aspect ratio and orientation were determined via the eigenvalues and the corresponding eigenvectors of $\Sigma$.}
\label{fig:Q1_SKH_B_perp}
\end{figure}

It is well known that an increasing out-of-plane field component $B_{\perp}$ leads to a shrinking skyrmion size and, above a certain threshold, to the transition of the skyrmionic spin structure to the collinear, field-polarized state. An additional change of the shape of the skyrmion must lead to a change of its parameters $\GYR$ and $\mathcal{D}$ with that  altering $\Theta(\varphi)$. This feature is illustrated in Fig. \ref{fig:Q1_SKH_B_perp} for the skyrmion with $\Q=-1$. We found that not only the size but also the aspect ratio of the topological skyrmion profile tends to increase with decreasing $B_{\perp}$, consequently leading to an increasing orientation dependence of the skyrmion Hall angle for larger $\Q=-1$ skyrmions. 

\begin{figure}
\includegraphics[scale=1.0]{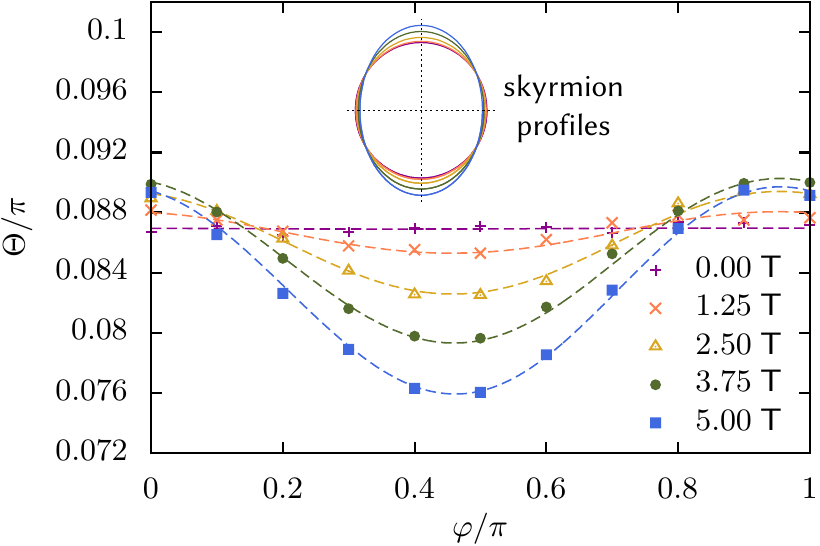}
\caption{Skyrmion Hall angle $\Theta$ versus direction of applied current $\varphi$ for skyrmions with $\Q=1$ 
%with $\alpha=0.1$ and $\beta=0.3$ 
for different strengths of the in-plane field $B_{\parallel}$. The orientation of the in-plane field is in $y$-direction, thus being parallel with the current for $\varphi=\pi/2$. Dashed lines correspond to our semi-analytical calculations following Eq.\ \ref{eq:Theta}. }
\label{fig:Q-1_SKH_B_inplane}
\end{figure}

However, varying the strength of $B_{\perp}$ cannot provoke an anisotropic dissipation tensor for skyrmions with $\Q=1$ and $\Q=-2$. Thus, we also applied an additional in-plane field in $y$-direction in order to deform the skyrmion profiles. The simulated $\Theta(\varphi)$ for otherwise isotropic $\Q=1$ skyrmions for different values of $B_{\parallel}$ are shown in Fig. \ref{fig:Q-1_SKH_B_inplane}. For increasing $B_{\parallel}$ we find indeed that the topological profile becomes increasingly elongated along the field direction, leading to a variation of the skyrmion Hall angle with $\varphi$. This effect can lead to both,
%, as predicted in Ref. \cite{Lin2015}.
an increase or a decrease of the skyrmion Hall angle $\Theta$, depending on $\varphi$.
% However, for the case of the $\Q=-2$ skyrmion we have observed that the shape is scarcely affected by an increasing $B_{\parallel}$ keeping its isotropic dissipation tensor up to a very high threshold above which it changes its shape significantly.

\section{Conclusion}
We examined the current-induced motion of metastable skyrmionic spin configurations with a variety of topological charges in a (Pt\textsubscript{0.95}Ir\textsubscript{0.05})/Fe bilayer on a Pd(111) surface by the means of spin model calculations. The effect of the spin-polarized current is modeled by supplementing the \ac{LLG} equation of motion with two additional torque terms, an adiabatic and a non-adiabatic spin-transfer-torque.

In agreement with previous findings, we find that skyrmionic structures propagate along trajectories which deviate from the current direction under the so-called skyrmion Hall angle which depends on the respective topological charge. Especially, the sign of this angle does depend on the sign of the topological charge. We have demonstrated that in this particular system the magnitude of the skyrmion Hall angle is in general not conserved when going from skyrmionic structures with $\Q = 1$ to topological charge $\Q=-1$. This effect is caused by the \ac{DMI} which renders skyrmions with $\Q = 1$ energetically favorable while deforming skyrmionic spin structures  with $\Q \neq 1$.

Moreover, due to the internal spin structure of skyrmionic structures, this deformation does not necessarily have to occur in an isotropic manner but instead can make skyrmions non-circular which consequently leads to an anisotropic response to a spin-polarized current. This was demonstrated by showing that the skyrmion Hall angle depends on the orientation of the skyrmion with respect to the current direction (varies up to $35 \%$).
Interestingly, this effect tends to increase with the size of the skyrmion (shown for $\Q=-1$). Moreover, we have shown that the rotational symmetry of the skyrmion with $\Q=1$ can be broken by applying an in-plane external field possibly opening up a new pathway to the manipulation of skyrmion dynamics.

Our numerical findings can be well explained within a generalized Thiele equation  \cite{Thiaville2005, Thiele1973}. Here, the current-driven skyrmion velocity depends on the topological charge as well as the shape dependent dissipation tensor. Usually is assumed the dissipation tensor can be replaced by a scalar quantity (e.g. Refs. \cite{Nagaosa2013,Iwasaki2013b}) which is justified for rotationally symmetric skyrmions. However, by dropping that assumption we derived an analytic expression for the skyrmion Hall angle that successfully describes its  dependence on the orientation of the skyrmions.

Our analysis of the topological density profiles of the skyrmionic structures leads to the conclusion that an non-circularity in shape does not necessarily lead to an orientation dependent Hall angle. Only a detailed, quantitative analysis of the properties of the respective dissipation tensor elements can lead to the understanding of the Skyrmion dynamics.

\begin{acknowledgments}
The authors thank the Deutsche Forschungsgemeinschaft (DFG, German Research Foundation) for financial support through the SFB 767 and project no. 403502522.

\end{acknowledgments}

\bibliographystyle{apsrev4-1}
\bibliography{bibfile}

\end{document}